\documentclass[a4paper,amsmath,amssymb,superscriptaddress,aps,prb,reprint,showpacs]{revtex4-2}
\usepackage{xcolor,graphicx}
\usepackage{subfigure}
\usepackage{MnSymbol}
\usepackage{braket}
\usepackage{bm}
\usepackage{mathrsfs}
\usepackage{tikz}
\usepackage{pgfplots}
\usepackage[utf8]{inputenc}

\newcommand{\ind}{_\mathrm}
\newcommand{\TN}{T_{N}}
\renewcommand{\i}{\mathrm i}
\newcommand{\cdag}{c^{\dagger}}
\newcommand{\cnod}{c^{\phantom{\dagger}}}
\newcommand{\e}{\textrm{e}}
\renewcommand{\vec}[1]{\mathbf{#1}}

\DeclareMathOperator*{\Tr}{Tr} 
\DeclareMathOperator*{\tr}{tr} 
\DeclareMathOperator*{\ii}{i} 
\DeclareMathOperator*{\ee}{e} 
\DeclareMathOperator*{\dd}{d} 
\newcommand{\mat}[1]{\bm{#1}} 

\begin{document}

\title{Suppression of effective spin-orbit coupling by thermal
  fluctuations in spin-orbit coupled antiferromagnets}

\author{Jan Lotze}
\affiliation{%
Institut f\"ur Funktionelle Materie und Quantentechnologien,
Universit\"at Stuttgart,
70550 Stuttgart,
Germany}
\author{Maria Daghofer}
\affiliation{%
Institut f\"ur Funktionelle Materie und Quantentechnologien,
Universit\"at Stuttgart,
70550 Stuttgart,
Germany}
\affiliation{Center for Integrated Quantum Science and Technology, University of Stuttgart,
Pfaffenwaldring 57, 
70550 Stuttgart, Germany}

\date{\today}

\begin{abstract}
We apply the finite-temperature variational cluster approach to a strongly
correlated  and spin-orbit coupled model for four electrons (i.e. two holes) in the
$t_{2g}$ subshell. We focus on parameters suitable for antiferromagnetic Mott
insulators, in particular Ca$_2$RuO$_4$, and identify a crossover from the
low-temperature regime, where spin-orbit coupling is essential, to the
high-temperature regime where it leaves few signatures. The crossover is seen in
one-particle spectra, where $xz$ and $yz$ spectra are almost one
dimensional (as expected for weak spin-orbit coupling) at high
temperature. At lower temperature, where spin-orbit coupling mixes all
three orbitals, they become more two dimensional. However, stronger
effects are seen in two-particle observables like the weight in states
with definite onsite angular momentum. We thus identify the enigmatic
intermediate-temperature 'orbital-order phase transition', which has
been reported in various X-ray diffraction and absorption experiments at
$T\approx 260\;K$, as the signature of the onset of
spin-orbital correlations.
\end{abstract}

\maketitle

\section{Introduction} \label{Sec:intro}
 Ruthenium oxides have for decades attracted considerable attention, first for
their complex phase diagrams that bear some similarities to those high-$\TN$
cuprate superconductors, with superconducting and Mott insulating
phases~\cite{RevModPhys.75.657}. More recently, the  interplay of spin-orbit
coupling (SOC), electron itineracy, electronic correlations, and lattice degrees of
freedom, which are all present, has attracted attention. The exotic behavior emerging on this stage includes a potential spin
liquid in
$\alpha$-RuCl$_3$~\cite{PhysRevB.90.041112,PhysRevB.91.094422},
enigmatic superconductivity in Sr$_2$RuO$_4$~\cite{Sr2RuO4_notriplet} and a non-equilibrium Weyl semi metal  in Ca$_2$RuO$_4$~\cite{Sow1084}.   

This last compound, Ca$_2$RuO$_4$, had already been
discussed as a Mott insulating end member of the family of compounds
including enigmatic superconductors.
Its high-temperature metal-insulator transition has been
well described by a combination of density-functional theory and dynamical
mean-field theory~\cite{PhysRevB.95.075145}. The emerging picture is that of a lattice-supported
Mott transition, where the $xy$ orbital is lowered in energy and
becomes nearly doubly occupied, while a Mott gap opens in the
approximately half filled $xz$ and $yz$ orbitals. 
SOC, which had alternatively been argued to drive the
metal-insulator transition\cite{PhysRevB.84.235136}, was later shown to
have only a weak impact on the gap~\cite{PhysRevB.95.075145}. 

However, this changes decisively when it comes to the magnetic
properties of the antiferromagnetic state observed at even lower
temperatures. For weak SOC and dominant crystal field (CF),
one would expect the half-filled $xz$ and $yz$ orbitals to form a spin
one, while the doubly occupied $xy$ orbital would be magnetically
inert. However, magnetic excitations turn out to show pronounced
$X$-$Y$-symmetry  as well as Higgs
modes~\cite{PhysRevLett.119.067201,Higgs_Ru}, which can more naturally
be explained in terms of 'excitonic' antiferromagnetism, which fundamentally relies
on SOC.

Orbital angular momentum of
two  $t_{2g}$ holes can be modeled as an effective  $L=1$. In the
idealized picture of an undistorted Ru-O octahedron (i.e. with
equivalent $xy$, $yz$, and $xz$ orbitals) SOC would
couple total spin $S=1$ with $L=1$ into a singlet ground state with
total angular momentum $J=0$~\cite{Khaliullin:2013du}. When superexchange connects ions, however, higher-energy triplets
gain kinetic energy and may condense into a magnetically ordered
state. In one dimension, the resulting ground-state phase diagram has
been established by use of the density-matrix
renormalization group and includes a parameter regime supporting
excitonic magnetism~\cite{PhysRevB.96.155111,PhysRevB.101.245147}.
Recent numerical work using a combination of density-functional theory
and variational cluster approach (VCA) has further indicated
that the excitonic scenario with SOC as a decisive player indeed
applies to the antiferromagnetic low-temperature state of
Ca$_2$RuO$_4$~\cite{PhysRevResearch.2.033201}.

It would thus be highly desirable to investigate temperatures between the
ground state with a large role for SOC and the high-temperature state,
where it only yields small corrections, also with a view towards other
ruthenates with similar energy scales. 
While the metal-insulator transition and the interplay of lattice and correlations is accessible
to  dynamical mean-field theory with a Monte-Carlo impurity
solver, the fermionic minus-sign problem is present at lower
temperatures~\cite{PhysRevLett.123.137204}. Adjusted one-particle
states based on total angular momentum can reduce
the minus-sign problem~\cite{PhysRevB.99.075117}, however, such an optimal
basis cannot easily be identified in realistic models, where CFs
or anisotropic hoppings compete with SOC. 

In order to address low temperature scales of spin-orbit coupled
$t_{2g}$ orbitals, described by the three-orbital Hubbard model of
Sec.~\ref{Sec:model}, we thus implement a finite-temperature variant
of the VCA, see Sec.~\ref{Sec:VCA}. Exact
diagonalization   is used to solve a small cluster and to extract
its self energy, which is then used to evaluate the Green's function of
the thermodynamic limit. This allows us to treat the antiferromagnetic
order, and since we focus here on the Mott insulating regime, bath sites are less necessary.

Based on the results presented in Sec.~\ref{Sec:results}, we identify a temperature range above the
N\'eel temperature, but in the Mott insulating regime, where the
spin-orbital character strongly changes. While the onsite-singlet and
triplet states describe the ionic state at low temperatures very well, as
expected for the excitonic scenario, they become less useful at higher temperatures. Here, 
the original orbitals provide a clearer picture, especially in the
presence of a CF, as will be seen in the one-particle
spectra discussed in Sec.~\ref{Sec:Akw}.

In $t_{2g}$ models
with SOC of a magnitude suitable for excitonic magnetism, there is
thus a third temperature scale intermediate between the
metal-insulator transition related to charge fluctuations and lattice
distortions and the N\'eel temperature related to magnetic degrees of
freedom. We discuss in Sec.~\ref{Sec:conclusions} how this  ties in with the
enigmatic 'orbital ordering' transition that has been debated at
intermediate temperatures in
Ca$_2$RuO$_4$~\cite{PhysRevLett.95.136401,PhysRevLett.87.077202}.

\section{Model and Methods} \label{Sec:model}

We study here a three-orbital Hubbard model for $t_{2g}$ electrons on a square
lattice, where we focus on nearest-neighbor (NN) hopping and tetragonal
symmetry.  The kinetic energy is then diagonal in orbital indices and takes the form 
\begin{align}\label{Hkin}
  H\ind{kin}&= -t \sum_{\langle i,j\rangle,\sigma}
  \cdag_{i,xy,\sigma}\cnod_{j,xy,\sigma}\\ \nonumber
  &\quad -t \sum_{\langle i,j\rangle\parallel x,\sigma}
  \cdag_{i,xz,\sigma}\cnod_{j,xz,\sigma}
  -t \sum_{\langle i,j\rangle\parallel y,\sigma}
  \cdag_{i,yz,\sigma}\cnod_{j,yz,\sigma}+ \textrm{H.c.}
\end{align}
where $\cnod_{i,\alpha,\sigma}$ ($\cdag_{i,\alpha,\sigma}$) annihilates
(creates) an electron with spin $\sigma=\uparrow,\downarrow$ in orbital
$\alpha=xy,xz,yz$ at site $i$. Nearest-neighbor bonds
$\langle i,j \rangle$ along the two direction $x$ and $y$ are
considered and we use $t$ as our unit of energy. 

Tetragonal CF splitting
\begin{align}\label{HCF}
  H\ind{CF}&= - \Delta \sum_{i,\sigma} n_{i,xy,\sigma}
\end{align}
with $n_{i,xy,\sigma} = \cdag_{i,xy,\sigma}\cnod_{i,xy,\sigma}$ and $\Delta > 0$ is motivated by the shortened octahedra of the
low-temperature phase of Ca$_2$RuO$_4$. For a filling of four
electrons (two holes), it favors a doubly occupied $xy$ orbital with
half filled $xz$ and $yz$ orbitals. We will tune it  to interpolate between an orbitally polarized spin-one system at large
$\Delta$ and a more equal interplay of degenerate orbitals at small $\Delta$.  

In the present paper, the impact of SOC is particularly important, which takes
the form 
\begin{equation}
  H\ind{SOC}=\lambda \sum_i \vec{l}_{i} \cdot \vec{s}_{i}
=
\frac{i\lambda}{2}\sum_i\sum_{\stackrel{\alpha,\beta,\gamma}{\sigma,\sigma'}} 
\varepsilon^{\phantom{\alpha}}_{\alpha\beta\gamma} \tau^\alpha_{\sigma\sigma'}
\cdag_{i,\beta,\sigma}\cnod_{i,\gamma,\sigma'}
  \label{HSOC}
\end{equation}
for $t_{2g}$ orbitals, with the totally antisymmetric
Levi-Civita tensor $\varepsilon_{\alpha\beta\gamma}$ and Pauli matrices
$\tau^\alpha$,  $\alpha=x,y,z$
\cite{PhysRevB.73.094428,PhysRevB.98.205128}. We focus here on intermediate
SOC that is not strong enough to suppress magnetic
ordering. 

Finally, there are effective onsite Coulomb interactions~\cite{PhysRevB.28.327} 
\begin{align} \label{Hint}
H\ind{int} &= U \sum_{i, \alpha} n_{i \alpha \uparrow} n_{i \alpha \downarrow} 
  +\frac{U^\prime}{2} \sum_{i, \sigma} \sum_{\alpha \neq \beta} n_{i \alpha
    \sigma} n_{i \beta \bar{\sigma}}\\ \nonumber
    & +\frac 1 2 (U^\prime - J_H) \sum_{i,\sigma} \sum_{\alpha \neq \beta}
  n_{i \alpha \sigma} n_{i \beta \sigma}\\ \nonumber
    & - J_H \sum_{i, \alpha \neq \beta} ( c^\dagger_{i \alpha \uparrow}
  \cnod_{i \alpha \downarrow} c^\dagger_{i, \beta \downarrow} \cnod_{i \beta
    \uparrow}- c^\dagger_{i \alpha \uparrow} c^\dagger_{i \alpha \downarrow} \cnod_{i \beta \downarrow} \cnod_{i \beta \uparrow})
\end{align}
with Coulomb interaction $U$, $U^\prime$ and Hund's coupling $J\ind H$
connected via $U^\prime = U-2J\ind H$. We are here less interested in
varying interactions and rather focus on the Mott insulating regime
with a Hund's-rule coupling larger than SOC, so that $\vec{L}$-$\vec{S}$ coupling provides a clearer description
than $j$-$j$ coupling. We thus choose $U=12.5t$ and $J_H=2.5t$, which is consistent with their order of
magnitude in Ca$_2$RuO$_4$~\cite{hund_Ca2RuO4}.

\subsection{Finite-temperature variational cluster approach} \label{Sec:VCA}

We use the finite-temperature~\cite{Yunoki2018PRB} 
VCA~\cite{PhysRevLett.91.206402,Pot03}, where the grand potential
$\Omega$ of
the system is approximated in terms of a 'reference system' that
consists of small disconnected clusters, but has the same
electron-electron interactions as the Hamiltonian of interest~\cite{Potthoff2012}: 
\begin{equation}
\Omega(\Sigma_{\text{cl}}) = \Omega_{\text{cl}} + \Tr\ln(-G_{\text{cl}}^{-1}) - \Tr\ln(-G_{\text{CPT}}^{-1})
\end{equation}
with the grand potential $\Omega_{\text{cl}}$ and Green's
function $G_{\text{cl}}$ obtained from the cluster. The CPT-Green's Function
$G^{-1}_{\text{CPT}}=(G_{\text{cl}}^{-1}-G_{\text{cl},0}^{-1}+G_{0}^{-1})^{-1}$
replaces the non-interacting cluster Green's function
$G_{\text{cl},0}$ by the
non-interacting Green's function $G_{0}$ of the
full system. The approximation thus consists in replacing the self
energy of the physical system by that of the small cluster. In order
to improve the approximation, the self-energy--functional approach~\cite{Pot03} allows us to
optimize the cluster self energy
$\Sigma_{\text{cl}}$ by varying one-particle parameters of the
reference Hamiltonian. The best approximation to the system's grand potential is a
stationary point of $\Omega$ w.r.t. the  variational one-particle
parameters.  Note that this variation affects only the small
cluster, the non-interacting Green's function
given by the one-particle part of the physical Hamiltonian remains
fixed. 

We numerically evaluate  the cluster grand potential
\begin{align}
  \Omega_{\text{cl}}=-\beta^{-1}\ln(\Xi)=-\beta^{-1} \ln \sum_{m}\e^{-\beta\varepsilon_{m}}
\end{align}
with partition function $\Xi$  and cluster-energies $\varepsilon_{n}$, as well as the cluster
Green's function, whose electron part reads 
\begin{equation}
[G_{\text{cl}}^{(+)}]_{\alpha\beta}(z) = \sum\limits_{m,n} \frac{\ee^{-\beta \varepsilon_{m}}}{\Xi} \left[\raisebox{-2mm}{$\dfrac{\Braket{\Psi_m|c_\alpha\vphantom{c_\beta^\dagger}|\Psi_n^{+}}\Braket{\Psi_n^{+}|c_\beta^\dagger|\Psi_m}}{z-E_{nm}^{+}+ \ii 0^{+}}$}\right] \label{eq:green}
\end{equation}
with energy difference $E_{nm}^{\pm}=\varepsilon_{n}^{\pm}-\varepsilon_{m}$.
 We largely follow
\citeauthor{Yunoki2018PRB}\cite{Yunoki2018PRB} and obtain the spectrum
and eigenstates with band Lanczos to resolve (approximately) degenerate eigenenergies. We use
eight starting vectors and converge 120 eigenvectors, which are used
to evaluate the Green's functions in a second Lanczos run. In this
second step, band Lanczos did not turn out to be advantageous, and we
thus use the conventional algorithm.

Assembling the cluster Green's function is accelerated with the help of a
high-frequency expansion for frequency
arguments with absolute values larger than that of the largest
pole~\cite{Yunoki2018PRB}. [Other frequencies are obtained via
\eqref{eq:green}.] Following \citeauthor{Yunoki2018PRB}, the
high-energy Green's function is expanded to $15$-th order as 
\begin{equation}
G_{\alpha\beta}(z) = \sum\limits_{k=0}^{\infty} \frac{M_{\alpha\beta}^{(k)}}{z^{k+1}}
\end{equation}
with moments $M_{\alpha\beta}^{(0)}(z)=\delta_{\alpha\beta}$ and
\begin{align}
M_{\alpha\beta}^{(k>0)} &= \sum\limits_{m,n} (E_{nm}^{+})^{k}\frac{\ee^{-\beta \varepsilon_{m}}}{\Xi} \Braket{\Psi_m|c_\alpha\vphantom{c_\beta^\dagger}|\Psi_n^{+}}\Braket{\Psi_n^{+}|c_\beta^\dagger|\Psi_m} \notag \\
&+ \sum\limits_{m,n} (-E_{nm}^{-})^{k} \frac{\ee^{-\beta \varepsilon_{m}}}{\Xi} \Braket{\Psi_m|c_\beta^\dagger|\Psi_n^{-}}\Braket{\Psi_n^{-}|c_\alpha\vphantom{c_\beta^\dagger}|\Psi_m}.
\end{align}

In order to fix the density to $N=4$ electrons (i.e. two 
holes), the grand potential is transformed to the free energy
$F(N,V,T)=\Omega(\mu,V,T)+\mu N$ by means of a Legendre
transform\cite{Balzer2010PRB}. There are thus at least two 
variational parameters, the chemical potential $\mu$ to fix the
density and the cluster chemical potential $\mu^{\prime}$ to ensure
thermodynamic consistency\cite{AichhornEPL72_117,Senechal2010}. Additionally, we
consider antiferromagnetic order parameters with ordered moment within
the $a$-$b$-plane or along the $c$-axis. Previous work for $T=0$ has shown that
the $z$ and in-plane components lead to quite different grand
potentials (as expected for finite SOC), but that the grand potential
is very similar for operators like spin, magnetization, or total
angular momentum~\cite{PhysRevResearch.2.033201}. For this reason and in
order to easily compare to the spin-one antiferromagnet, we use here
the spin as the order parameter.  It has also been shown
that a sizable CF as well as SOC both favor 'checkerboard'
magnetic patterns with ordering vector $\vec{Q}=(\pi,\pi)$, so that we
use the fictitious Weiss field
\begin{align}\label{eq:weiss}
  H_{\textrm{Weiss}} = h \sum_{i} \e^{\i \vec{Q}\vec{r}_i} S_i^{x/z}
\end{align}
with $i$ labeling the site at $\vec{r}_i$ and $S_i^{x}=\sum_{\alpha}
\cdag_{i,\alpha,\uparrow}\cnod_{i,\alpha,\downarrow} + \textrm{H.c.}$
and
$S_i^{z}=\sum_{\alpha}
(\cdag_{i,\alpha,\uparrow}\cnod_{i,\alpha,\uparrow}- \cdag_{i,\alpha,\downarrow}\cnod_{i,\alpha,\downarrow})$
are the $x$ and $z$ component of the total spin, respectively. 

For variational parameters $\mu$, $\mu'$ and $h$ giving stationary
grand potentials, we evaluate one-body expectation values (like
magnetization or orbital densities) from the
Green's function, as is done for $T=0$. The entropy $S$ is  determined
as the derivative of the grand potential via
contour integrals: 
\begin{equation}
S = S_{\text{cl}} + \mathscr{S}_{\text{CPT}} + \mathscr{S}_{\text{cl}} - (\Omega_{\text{CPT}}-\Omega_{\text{cl}})/T
\end{equation}
with the contributions
\begin{align}
S_{\text{cl}} &= -(\Omega_{\text{cl}}-\braket{H}_{\text{cl}})/T,\\
\mathscr{S}_{\text{CPT}} &= \oint_{\mathcal{C}}\dd z f(z) \tr[ (\mat{G}_{\text{cl}}\mat{G}_{\text{CPT}}^{-1}\mat{G}_{\text{cl}})^{-1} \mat{G}_{\text{mod}}]/T^{2},\\
\mathscr{S}_{\text{cl}} &= \oint_{\mathcal{C}}\dd z f(z) \tr[ (-\mat{G}_{\text{cl}}^{-1}) \mat{G}_{\text{mod}}]/T^{2}
\end{align}
and the abbreviations
\begin{align}
\braket{H}_{\text{cl}} &= \Xi^{-1}\sum_{m}\varepsilon_{m}\exp(-\beta\varepsilon_{m}),\\
\mat{G}_{\text{mod}} &= \sum_{m,n}(\braket{H}_{\text{cl}}-\varepsilon_{m})\Xi^{-1}\exp(-\beta\epsilon_{m})[\mat{G}^{+}_{\text{cl}}+\mat{G}^{-}_{\text{cl}}] \notag \\
&-\sum_{m,n}(T z)\Xi^{-1}\exp(-\beta\epsilon_{m})\left[\frac{\partial\mat{G}^{+}_{\text{cl}}}{\partial z}+\frac{\partial\mat{G}^{-}_{\text{cl}}}{\partial z}\right].
\end{align}
The specific heat $C(T)$
is then obtained as the numerical derivative $C(T)=T(\Delta S/\Delta T)$ of the entropy. 

Since  the ordered moment in an excitonic magnet arises through a superposition of
the $J=0$ state preferred by onsite SOC and $J=1$
states~\cite{Khaliullin:2013du}, it is helpful to consider weights
$\langle J \rangle$ found in eigenstates of total onsite angular
momentum $J$. Unfortunately, this is a two-particle
quantity and thus not readily available from the VCA. We approximate it
as the exact-diagonalization expectation value obtained for the $N_{\textrm{sites}}$
sites of the reference cluster with optimized
parameters:
\begin{align}\label{eq:wght_j}
\langle J \rangle := \frac{1}{N_{\textrm{sites}} \Xi}
\sum_{i,J^z}\sum_m \e^{-\beta \varepsilon_m}\langle m| J_i,J^z_i \rangle \langle J_i,J^z_{i}
|m\rangle\;.
\end{align}
Eigenstates $|J_i,J_i^z\rangle$ denote here the state at site $i$ defined by
angular momentum $J=L+S$.

\section{Results} \label{Sec:results}  

\subsection{Temperature dependence of the onsite angular momentum}\label{Sec:J_T}

\begin{figure}
  \includegraphics[width=\columnwidth]{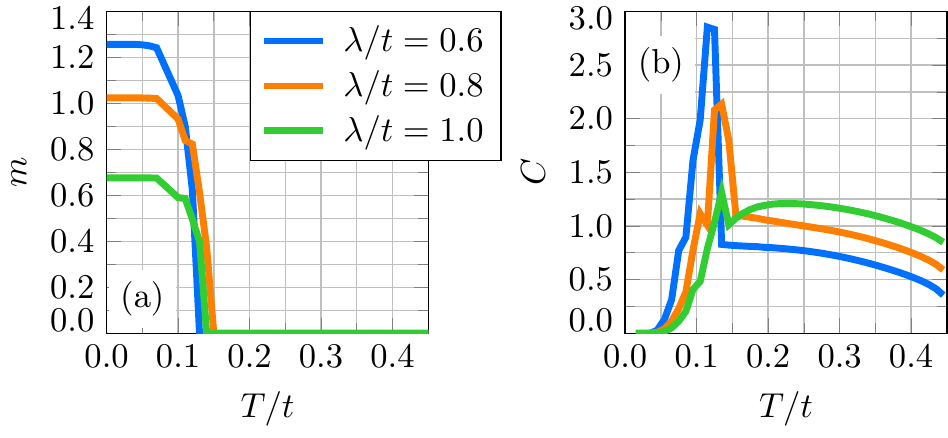}
  \caption{Thermodynamics of the excitonic antiferromagnet without CF splitting. (a) gives the
    magnetic order parameter, i.e. staggered out-of-plane spin and (b) the
    specific heat for SOC $\lambda =0.6 t, 0.8 t, t$. (We do not intend to discuss the complicated spin and orbital
    stripe pattern found at  $\lambda = \Delta=
    0$~\cite{PhysRevResearch.2.033201}, and accordingly leave out the
    regime of small $\lambda \lesssim 0.4 t$.)  \label{fig:thermo_CF0}}
\end{figure}

At $T=0$, VCA for the $t_{2g}$ model with four electrons and CF
$\Delta=0$ has revealed two different
ordering patterns depending on $\lambda$~\cite{PhysRevResearch.2.033201}: at small $\lambda \lesssim
0.4\;t$, orbitals and spins order in a stripy pattern with orthogonal
ordering momenta $(0,\pi)$ for spins and $(\pi,0)$ for orbitals. For
larger $\lambda$, excitonic antiferromagnetic (AFM) order with momentum $(\pi,\pi)$ takes
over, where the out-of-plane $z$ component is favored over in-plane
directions. We are here interested in the latter regime and thus focus on $\lambda > 0.4\;
t$.  

Figure~\ref{fig:thermo_CF0}(a) shows the ordered spin moment depending on
temperature. As expected, the value at $T=0$ is reduced when larger
$\lambda$ increases the energy gap between the ionic singlet and
triplet states, which in turn reduces the triplet admixture into the
ground state. Somewhat surprisingly, the N\'eel temperature is not
monotonic. While we can certainly not exclude strong finite-size
effects due to the $2\times 1$-site cluster, an alternative explanation may
be that system at smallest $\lambda =0.6\;t$ is affected by its
closeness to the competing stripy phase. The corresponding specific
heat is given in Fig.~\ref{fig:thermo_CF0}(b) and has a second broad hump at higher temperature $T > \TN$ in
addition to the expected peak at the magnetic ordering
transition. This feature exists for all three values of $\lambda$ and
shifts to slightly higher temperatures for $\lambda=t$. 

\begin{figure}
  \includegraphics[width=\columnwidth]{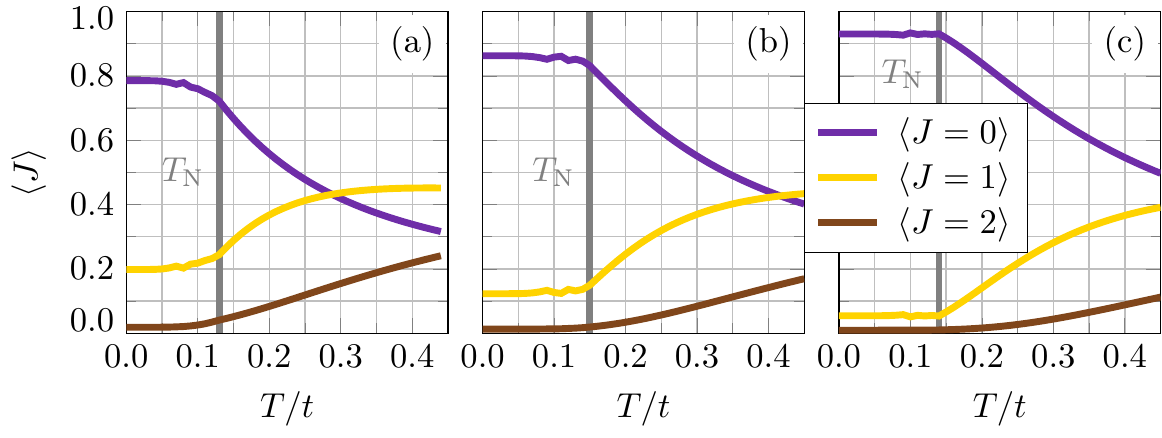}\\
  \caption{Temperature evolution of the average weight found in eigenstates of the total angular
    momentum, see Eq.~(\ref{eq:wght_j}).  (a) $\lambda=0.6 t$, (b) $\lambda =0.8 t$, and (c) for 
    $\lambda=t$.  \label{fig:J012_CF0}}
\end{figure}

Figure~\ref{fig:J012_CF0} shows the average weight
Eq.~(\ref{eq:wght_j}) found in eigenstates with $J=0,\;1,\;2$ of the
total onsite angular momentum. Weights are constant in the
regime of constant magnetization, and the $J=0$ ($J=1$) state looses
(gains) weight when magnetic order is lost. This is in clear contrast
to a (somewhat artificial) transition to a paramagnet at constant temperature:
reducing the ordering field $h$ at $T=0$ pushes weight from the $J=1$
states into the $J=0$ state~\cite{PhysRevResearch.2.033201}. At $\TN$, the
curves get abruptly steeper and weights in $J=0$ and $J=1$ states
change substantially at higher temperatures $T > \TN$. Weight in the $J=2$
states is completely negligible below $\TN$, but similarly begins to
grow at $T> \TN$. We are going to argue that this spin-orbital
rearrangement is the origin of the second hump in the specific heat. 

\begin{figure}
  \includegraphics[width=\columnwidth]{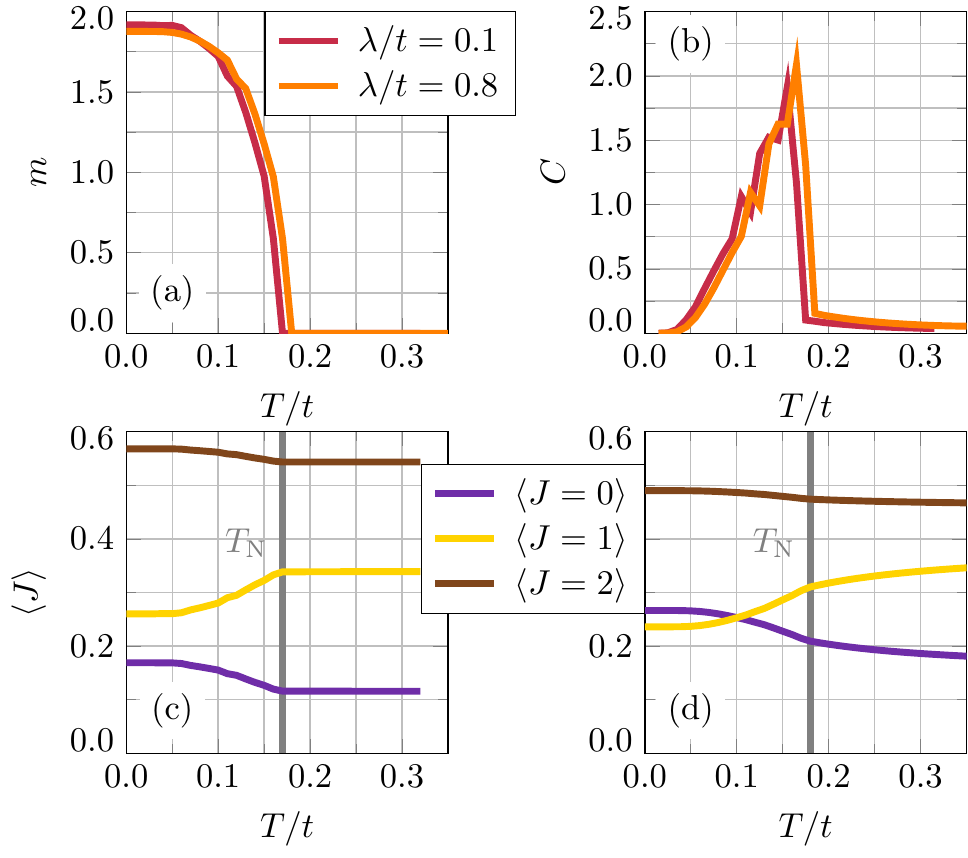}
  \caption{Thermodynamics of spin antiferromagnet at large CF $\Delta
    = 5t$. (a) gives the
    magnetic order parameter, i.e. staggered magnetization and (b) the
    specific heat for SOC $\lambda =0.1 t$ and $\lambda =0.8t$. ($\lambda
    =0$ was numerically less stable.) (c) and (d) give the average
    overlaps of Eq.~(\ref{eq:wght_j}) for $\lambda =0.1 t$ and $\lambda
    =0.8t$, resp. 
    \label{fig:thermo_CF5}}
\end{figure}

For comparison, Fig.~\ref{fig:thermo_CF5}(a) gives the
magnetization and specific heat for CF $\Delta = 5\;t$  that is large
enough to enforce complete orbital polarization with a doubly occupied
$xy$ orbital at all temperatures shown. The two holes are then found
in $xz$ and $yz$ orbitals and form a conventional spin one, with an
ordered moment close to two in the AFM state. The specific heat shown
in Fig.~\ref{fig:thermo_CF5}(b) has here only the peak at the N\'eel
temperature and no further features. The expected weights in eigenstates
with total onsite angular momentum $J=0,1,2$ are given in
Fig.~\ref{fig:thermo_CF5}(c) and (d) and present a quite different
picture from the  excitonic case discussed in
Fig.~\ref{fig:J012_CF0}: while the weights in $J=0$ and $J=1$ states
change appreciably below $\TN$, only little variation
is seen above $\TN$. 

\begin{figure}
  \includegraphics[width=\columnwidth]{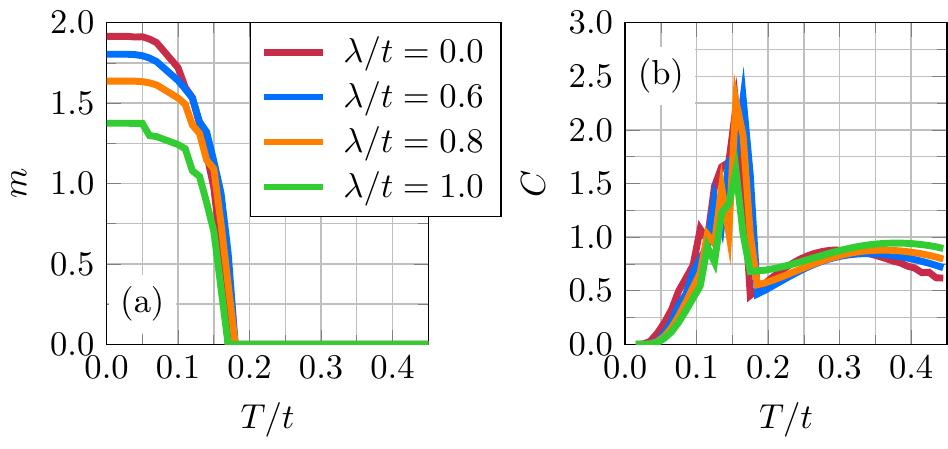}
  \caption{Transition from spin to excitonic antiferromagnet at
    $\Delta=1.5 t$. (a) gives the
    magnetic order parameter, i.e. staggered in-plane spin magnetization and (b) the
    specific heat for SOC $\lambda =0,\; 0.6 t, 0.8 t, t$.
  \label{fig:magn_C_D15}}
\end{figure}

\begin{figure}
  \includegraphics[width=\columnwidth,clip]{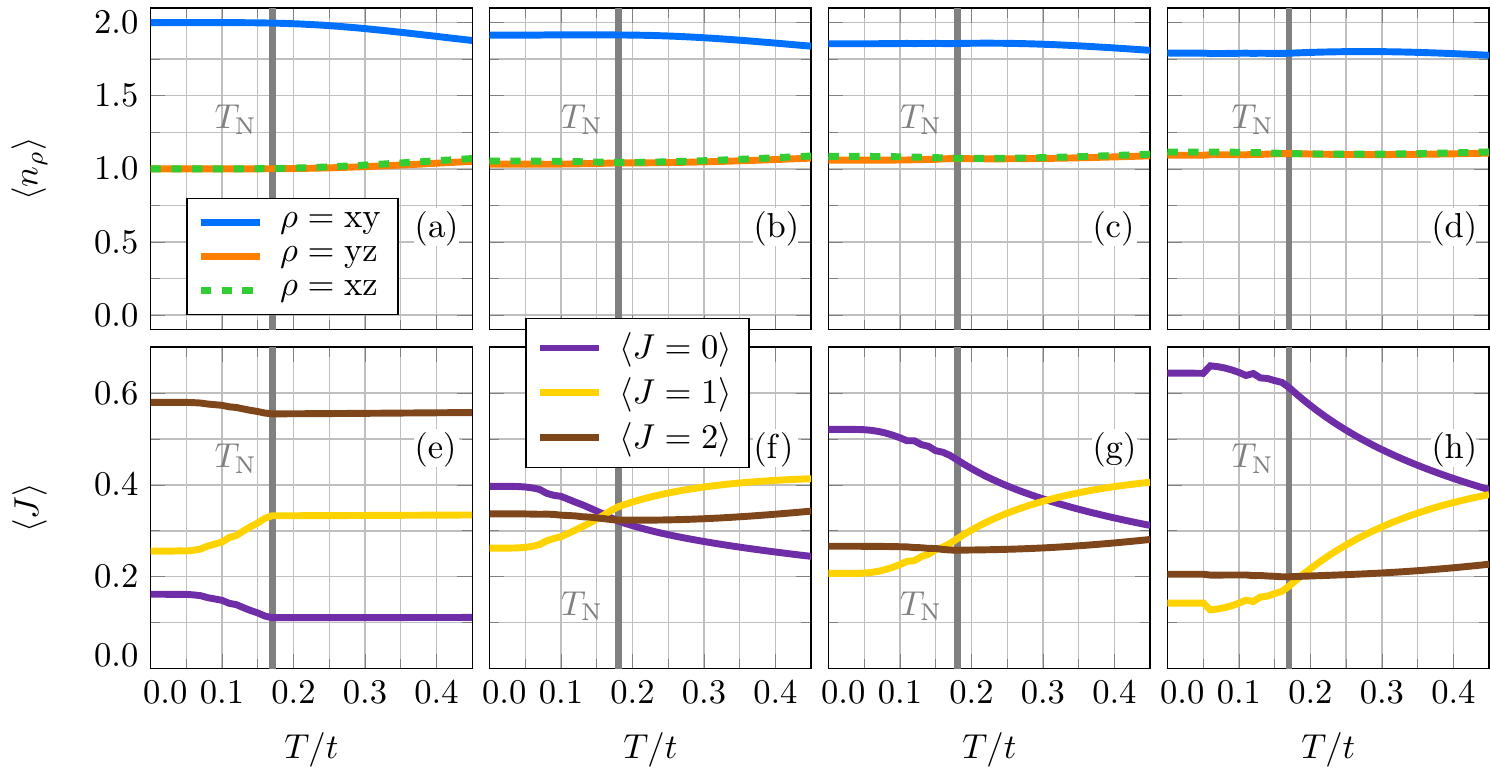}
  \caption{Spin-orbital onsite wave function for $\Delta=1.5 t$.
    (a-d) show   the orbital-resolved densities for $\lambda
    =0, 0.6 t, 0.8 t, t$ and (e-h) the weights in $J$ states according to
    Eq.~(\ref{eq:wght_j}). \label{fig:J_orb_15}}
\end{figure}

Finally, Figs.~\ref{fig:magn_C_D15} and ~\ref{fig:J_orb_15} discuss intermediate $\Delta = 1.5
t$, an order of magnitude appropriate to describe
Ca$_2$RuO$_4$. Ground-state VCA calculations have here shown in-plane
magnetic moments to be favored over out-of-plane moments~\cite{PhysRevResearch.2.033201}, in
agreement with the AFM state of Ca$_2$RuO$_4$.
Again, the N\'eel temperature is not very sensitive to
SOC $\lambda$ while the ordered moment is substantially reduced by
it.  The system without  SOC has the largest ordered moment close to
two, see Fig.~\ref{fig:magn_C_D15}(a). As will be discussed below, its 
$xy$ orbital is completely filled  below the N\'eel
temperature, see Fig.~\ref{fig:J_orb_15}(a), so that it comes close to
a spin-one scenario.
Larger $\lambda \geq 0.6\;t$ reduce the ordered
moment, which indicates that orbital polarization is not strong enough
to quench SOC.
The specific heat shown in Fig.~\ref{fig:magn_C_D15}(b)
looks qualitatively much more similar to the results  for $\Delta=0$
than to those for $\Delta=5\;t$, as a second hump at $T> \TN$ is
clearly seen.

While the transition from spin-one to excitonic
antiferromagnetism is a gradual crossover, it was estimated to occur
at $\lambda \approx 0.7 t$ in the ground
state~\cite{PhysRevResearch.2.033201}. The $J$-weights qualitatively agree,
with Fig.~\ref{fig:J_orb_15}(e) for $\lambda =0$ being similar to the spin-one scenario
of Fig.~\ref{fig:thermo_CF5}(c,d), while Figs.~\ref{fig:J_orb_15} (g,h) for $\lambda =0.8 t$ and
$\lambda=1$ resemble more the excitonic case of
Fig.~\ref{fig:J012_CF0}. Figure~\ref{fig:J_orb_15}(f) for $\lambda
=0.6 t$ lies  somewhere in between, again in line with the previous estimate. 

The second hump in the specific heat for $\lambda=0$ can be understood
by noting that the orbital densities in Fig.~\ref{fig:J_orb_15}(a) do not remain constant above
$\TN$. Since the CF is here just strong enough to fill the $xy$
orbital at $T=0$, finite temperature can induce $xy$-holes and these
orbital fluctuations are reflected in the specific heat. The weights
found in states $J=0,1,2$, in contrast do here not change above
$\TN$, see Fig.~\ref{fig:J_orb_15}(e), when there is no SOC.

In the opposite limit $\lambda =t$, the orbital densities are nearly
constant, a small difference between $xz$ and $yz$ below $\TN$ being
due to magnetic symmetry breaking. Weights in $J=0$ and $J=1$ states
depend here strongly on temperature at $T> \TN$, see
Fig.~\ref{fig:J_orb_15}(h). While low $T\lesssim \TN$ strongly
suppressed any weight in $J=2$ states for $\Delta=0$, see
Fig.~\ref{fig:J012_CF0}, it is here nearly constant, because it is
connected to  the clear  orbital polarization $n_{xy} >
n_{xz/yz}$~\cite{PhysRevResearch.2.033201}.  

\subsection{Signatures of SOC in one-particle spectra} \label{Sec:Akw}

\begin{figure}
  \includegraphics[width=\columnwidth]{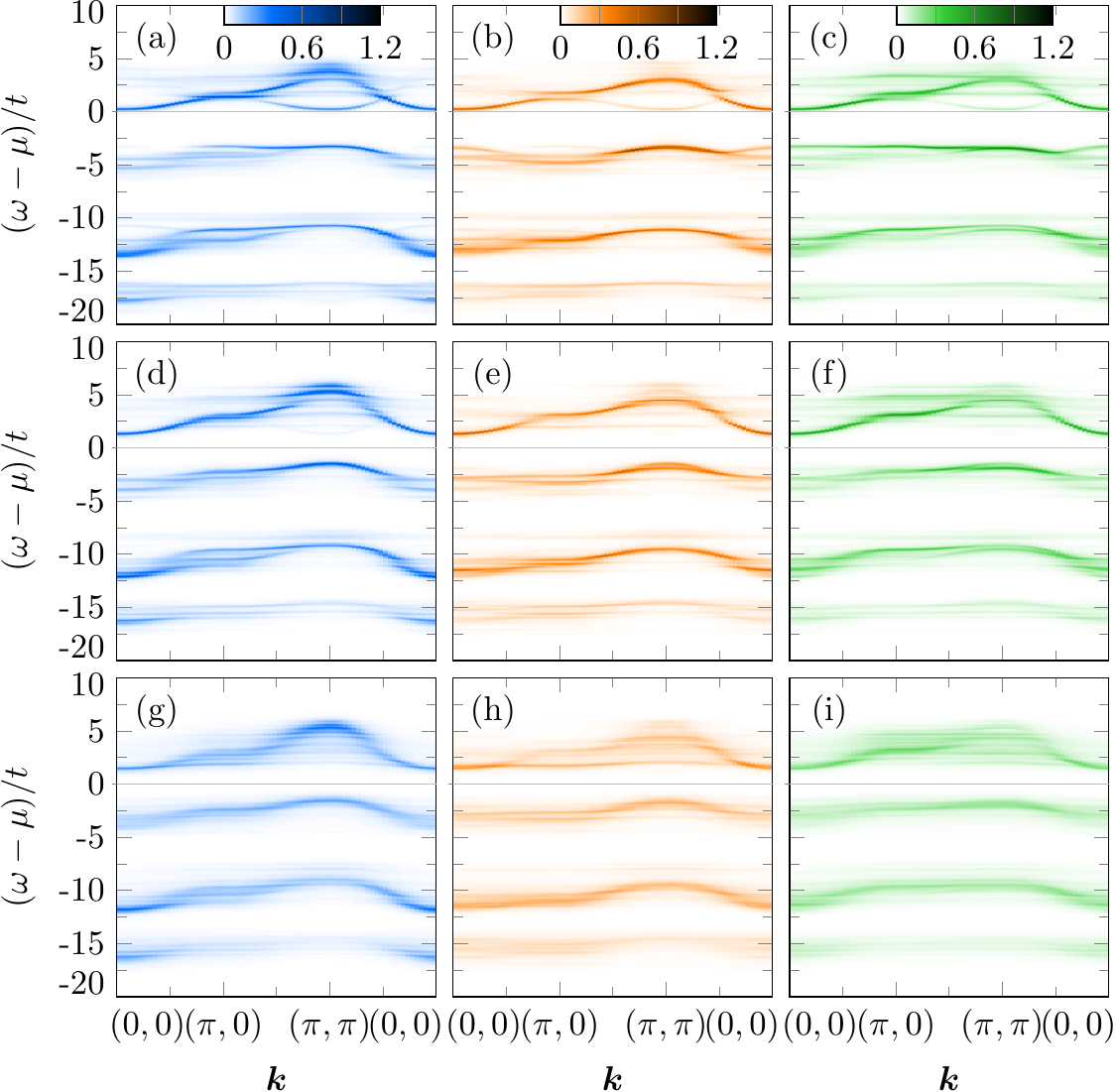}
  \caption{Orbital-resolved one-particle spectral density for  $\Delta=0$,
    $\lambda = 0.8 t$ and temperatures (a-c) $T=0$, (d-f) $T = 0.14 t \approx \TN$,
    and (g-i) $T= 0.35 t$. Orbital character is $xy$ in (a), (d), and (g),
    $yz$ in (b), (e), and (h), and $xz$ in (c), (f), and (i). \label{fig:Akw_Delta_0}} 
\end{figure}

Figure~\ref{fig:Akw_Delta_0} shows the VCA one-particle spectral
density for $\Delta=0$ and temperatures $T=0$, $T\gtrsim \TN$ and $T
\gg \TN$. At all temperatures, the occupied states are split into
three subbands at energies $\omega \lesssim 5 t$, $\omega \approx 10
t$, and $\omega \gtrsim 15 t$ (with the last having lower
weight), which can be related to Hund's-rule
coupling~\cite{hund_Ca2RuO4}.  Below $\TN$, some signatures of
the doubling of the unit cell are visible in the form of shadow bands
around $(0,0)$ and $(\pi,\pi)$. Apart from this feature, the
predominant effect of temperature is making the spectra less
coherent. Overall, temperature effects are here rather
subtle.

\begin{figure}
  \includegraphics[width=\columnwidth]{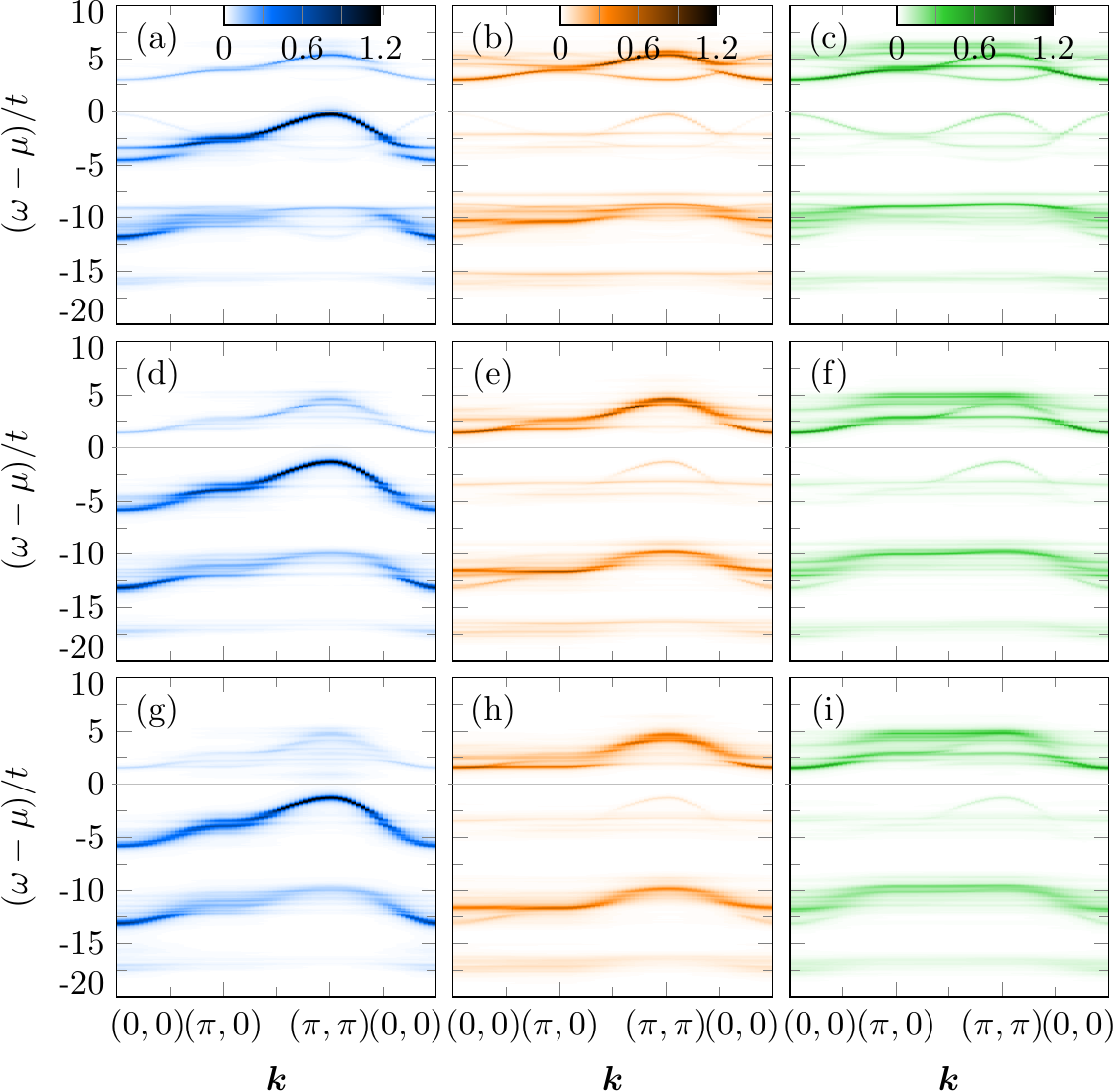}
  \caption{Orbital-resolved one-particle spectral density for  $\Delta=1.5 t$,
    $\lambda = t$ and temperatures (a-c) $T=0$, (d-f) $T = 0.16 t \approx \TN$,
    and (g-i) $T= 0.35 t$. Orbital character is $xy$ in (a), (d), and (g),
    $yz$ in (b), (e), and (h), and $xz$ in (c), (f), and (i). \label{fig:Akw_Delta_15}} 
\end{figure}

Temperature-driven orbital reconstruction reveals itself slightly more
when CF and SOC compete, see the one-particle spectra
shown in Fig.~\ref{fig:Akw_Delta_15} for $\Delta=1.5 t$ and $\lambda =t$. The
ground-state spectrum Fig.~\ref{fig:Akw_Delta_15}(a-c) shows again a slight shadow
band due to the doubling of the unit cell and both the empty band (of
predominantly $xz$/$yz$ character) and the highest occupied band (of predominantly $xy$ character)
have a two-dimensional dispersion, similar to Fig.~\ref{fig:Akw_Delta_0}(a-c). In the spectra taken around $\TN$, see
Fig.~\ref{fig:Akw_Delta_15}(d-f), the shadow band has vanished. The occupied $xz$/$yz$ states have
become more coherent than in the ground state. This rather
unconventional behavior may be related to the ladder-like features
that were recently found in a strong-coupling $t$-$J$-like model
without SOC, where they arise in the  AFM state due to the anisotropic
hoppings of these orbitals~\cite{PhysRevB.101.035115}: when magnetic order is
lost, the  ladder features become weaker and the underlying dispersion
is seen more easily. It is  rather one-dimensional, as expected for
$xz$/$yz$ orbitals without SOC. Such a weaker impact of SOC at higher
binding energies is somewhat reminiscent of the correlation-induced
energy dependence of SOC previously reported for 
metallic Sr$_2$RuO$_4$~\cite{PhysRevLett.120.126401}.

In the \emph{unoccupied} $xz$ and $yz$ states, on the other hand,
incoherent features have gained weight in addition to the
coherent band dominating the $T=0$ spectrum.  They do not follow the
two-dimensional dispersion of the coherent band,
but are more one dimensional. At high temperature $T=0.35 t$, finally, the unoccupied bands show
mostly the one-dimensional dispersion characteristic of $xz$/$yz$ orbitals in the absence
of SOC, see Fig.~\ref{fig:Akw_Delta_15}(h-i). In the
presence of a CF, SOC thus only
couples the three orbitals into a 2D dispersion at lower temperatures
and lower excitation energies, while spectra at higher temperatures
and energies look similar to the case without SOC.  

\section{Discussion and Conclusions} \label{Sec:conclusions}  

We have shown that temperature strongly affects the spin-orbital
onsite state in the PM Mott insulating state of spin-orbit coupled
$t_{2g}^4$ systems. We have investigated parameter sets supporting
excitonic AFM order at low temperatures, with and without a crystal
field. As long as the CF is not strong enough to completely
quench the orbital degree of freedom, we consistently find a second
broad hump in the specific heat, in addition to the peak at
$\TN$. In the same temperature range, onsite total angular momentum
changes substantially. 

In one-particle spectra, low-energy excitations stemming from $xz$
and $yz$ orbitals are two-dimensional in the ground state, but become
more one-dimensional at higher temperatures. This can also be
interpreted as SOC being most effective at low temperatures. Overall, signatures
of SOC and of the temperature-driven spin-orbital
rearrangement are rather subtle in one-particle spectra. Even at low
temperatures, where SOC is essential do reproduce the dispersion of
magnetic excitations~\cite{PhysRevLett.119.067201,Higgs_Ru}, one-particle spectra have thus been reasonably
well described already without taking SOC into
account~\cite{hund_Ca2RuO4,PhysRevB.101.035115}.  

However, we argue that X-ray diffraction and absorption experiments performed on
Ca$_2$RuO$_4$ show signatures of the spin-orbital rearrangement found here. Parameters for this compound correspond roughly to
those of Figs.~\ref{fig:magn_C_D15}(c,d) and~\ref{fig:Akw_Delta_15},
i.e. $\Delta\approx 1.5 t$ and $\lambda \approx 0.8 t - 1
t$~\cite{PhysRevResearch.2.033201}.
At temperatures of $\approx 260\;\textrm{K}$, i.e. between the
metal-insulator transition (which goes together with a structural
phase transition) and the N\'eel transition, signatures of another phase
transition were reported early on and interpreted in terms of orbital
order~\cite{PhysRevLett.95.136401,PhysRevLett.87.077202}.

Since this additional transition does not break any spatial
symmetries, one can rule out orbital stripe~\cite{PhysRevB.74.195124} or
checkerboard~\cite{PhysRevLett.88.017201} patterns theoretically
predicted for absent (or weak) SOC. More recent work established that
the transition cannot be related to a change in orbital
densities, leaving only the phase in a complex orbital
superposition as a possibility~\cite{PhysRevB.98.125142}. This would
fit with our findings of an SOC-driven spin-orbital
rearrangement. When SOC prefers the $J=0$ state at low temperatures,
this implies for each spin projection a specific phase relation
between the orbitals. In contrast, no definite phases are expected at higher
temperatures where SOC is less active. 

We have thus identified the enigmatic orbital-order transition in
Ca$_2$RuO$_4$ as a transition to a spin-orbit coupled onsite wave
function. This implies, e.g.,  that a spin up (down) prefers the complex
$|l^z=\pm 1\rangle$ orbital over the opposite state. This is somewhat 
reminiscent of ferro-orbital order into complex orbitals, which was early on
proposed as a scenario for
Ca$_2$RuO$_4$~\cite{PhysRevLett.87.077202}. More generally, complex-orbital order has been suggested to play a role
in doped manganites~\cite{PhysRevB.63.140416} and the Vervey
transition of magnetite~\cite{doi:10.1143/JPSJ.77.074711}. Spontaneous
complex-orbital order is, however, rare, because lattice distortions
like the Jahn-Teller effect favor real orbitals. The present
work not only reconciles this picture with the observation of (nearly)
constant density on $xy$ orbitals in Ca$_2$RuO$_4$, but moreover shows the transition
to arise naturally in a three-orbital model with SOC. 

\begin{acknowledgments}
The authors acknowledge support by the state of Baden-W\"urttemberg through
bwHPC and via the Center for Integrated Quantum Science and Technology
(IQST). This research was supported by the Deutsche Forschungsgemeinschaft
via FOR1807 (DA 1235/5-1).
\end{acknowledgments}

\bibliography{d5_system,d4_system,sonstige,bibliography,notes}

\end{document}